\documentclass[%
 reprint,
groupedaddress,
nobibnotes,
 amsmath,amssymb,
 aps,
prb,
]{revtex4-2}

\usepackage{amsthm,amsmath,amssymb}
\usepackage{mathrsfs}
\usepackage{caption}
\usepackage{graphicx, subfig}
\usepackage{bm}
\usepackage{float}
\usepackage[section]{placeins}

\begin{document}
	\preprint{APS/123-QED}
	
	\title{High magnetic field phase diagram and weak ferromagnetic
		moment breaking in (Ni$_{0.93}$Co$_{0.07}$)$_{3}$V$_{2}$O$_{8}$}

	
	\author{Jiating Wu$ ^{1,2} $, Minjie Zhang$ ^{1,2} $, Ke Shi$ ^{1,2} $, Huxin Yin$ ^{1,3} $, Yuyan Han$ ^{1} $, Lansheng Ling$ ^{1} $, Wei Tong$ ^{1} $, Chuanying Xi$ ^{1} $, Li Pi$ ^{1} $, Zhaosheng Wang$ ^{1,} $}
	\email{zswang@hmfl.ac.cn}
	\affiliation{$ ^{1} $Anhui Key Laboratory of Low-Energy Quantum Materials and Devices , High Magnetic Field Laboratory, HFIPS, Chinese Academy of Sciences, Hefei, Anhui 230031, China\\$ ^{2} $ University of Science and Technology of China, Hefei 230026, Anhui, China\\$ ^{3} $Engineering Technology Research Center of Magnetic Materials, School of Physics and Materials Science, Anhui University, Hefei 230601, China
	}%
	
	
	

	

	
	\date{\today}
	\begin{abstract}
		We present magnetostriction and thermal expansion measurements on multiferroic (Ni$_{0.93}$Co$_{0.07}$)$_{3}$V$_{2}$O$_{8}$. The high field phase diagrams up to 33 T along the $ a $, $ b $ and $ c $ directions are built. For H//$ a $, as the magnetic field increasing, two intermediate phases appear between the incommensurate phase and the paramagnetic phase at about 7 K, and then a magnetically induced phase appears above the paramagnetic phase. For H//$ b $, a thermal expansion measurement indicates a mutation in the spin lattice coupling of the high field phases. The interlaced phase boundary suggests a mixed state in the optical high field phase. For H//$ c $, an intermediate phase between the commensurate phase and the incommensurate phase is detected. A nonlinear boundary between the intermediate phase and the low temperature incommensurate phase, and a clear boundary between the commensurate phase and the paramagnetic phase are found. These results indicate that doping Co$^{2+}$ breaks the weak ferromagnetic moment of the commensurate phase, which exists in the parent compound Ni$_{3}$V$_{2}$O$_{8}$ and (Ni$_{0.9}$Co$_{0.1}$)$_{3}$V$_{2}$O$_{8}$. This nonlinear influence reflects complicated spin modulation in Ni$_{3}$V$_{2}$O$_{8}$ by doping Co$^{2+}$. 
	\end{abstract}
	\maketitle
	Magnetoelectric multiferroic materials raise attention due to the coexistence of magnetic and electric order, in which the type-\uppercase\expandafter{\romannumeral2} multiferroic materials could couple the ferromagnetism and ferroelectricity  \cite{spaldin_2017,doi:10.1126/science.1113357,PhysRevB.84.060406}. A number of type-\uppercase\expandafter{\romannumeral2} multiferroic materials were investigated in the past few decades, such as TbMnO$_{3}$, Fe$_{2}$(MoO$_{4}$)$_{3}$, MnWO$_{4}$ and Ni$_{3}$V$_{2}$O$_{8}$(NVO) \cite{PhysRevLett.98.147204,PhysRevLett.127.097601,PhysRevMaterials.6.094412,PhysRevLett.97.097203,PhysRevB.104.014415,PhysRevLett.93.247201}. NVO is a representative quantum spin system with a geometric frustrated kagome staircase, which has been widely reported\cite{PhysRevB.73.184433,PhysRevB.75.064427,PhysRevB.75.012407}. It is formed from corner sharing triangles with two kinds of nonequivalent $ S $=1 Ni$^{2+}$ spins in the spine and cross-tie sites\cite{PhysRevB.74.014429}. The geometric frustrated kagome staircase leads to unusual low temperature spin order and competing magnetic phases\cite{PhysRevLett.100.097202,PhysRevB.87.144404,PhysRevLett.84.2953}. Work on the NVO system shows complicated ground states led by spin correlation in magnetic fields\cite{PhysRevB.74.014429}. These ground states are susceptible to perturbations. As the temperature decreases,   the NVO system undergoes a paramagnetic(P) phase into a high temperature incommensurate(HTI) phase, low temperature incommensurate(LTI) phase, antiferromagnetic commensurate(C) phase and C$^{'}$ phase at T$_{PH}$(9.1 K), T$_{HL}$(6.3 K), T$_{LC}$(4 K) and T$_{CC'}$(2.3 K), respectively. In the LTI phase, the spontaneous electric polarization induced by a particular magnetic order in the spine and cross-tie sites has been found\cite{PhysRevLett.95.087205}. The complex boundaries of the magnetic phase diagram could be understood qualitatively by some simple models\cite{PhysRevB.74.014429}. For example, the sequence of phase transitions with decreasing temperature  is determined by Heisenberg nearest-neighbors(NN) and next nearest-neighbors(NNN) exchange interactions. The anisotropy K modulates the range of the HTI phase. The Dzyaloshinskii Moriya(DM)\cite{PhysRevB.78.140405} interaction and pseudodipolar(PD)\cite{PhysRevB.59.1079} interaction generate a weak ferromagnetic moment, leading to the linear boundary in HTI-C and LTI-C transitions while H//$ c $\cite{PhysRevB.74.014429}. For H//$ a $, as the fields increase, the high field phases arise, in which the F1 and F2 phases are ferroelectric and the F3 phase is paraelectric\cite{PhysRevB.84.220407}. In addition, another isomorphic compound, Co$_{3}$V$_{2}$O$_{8}$(CVO), exhibits distinct behavior. This compound is formed from S=$\dfrac{3}{2}$ Co$^{2+}$ spins with larger anisotropy than Ni$^{2+}$\cite{PhysRevB.74.014430}. The magnetic phase diagram of this compound is relatively simple and has no ferroelectricity\cite{doi:10.1143/JPSJ.76.034706}. It is worth mentioning that doping 10$\%$ Co$^{2+}$ on NVO can suppress the C phase and improve the stabilization of the multiferroic cycloidal phase LTI while H//a\cite{PhysRevB.94.174441}. Furthermore, the (Ni$_{0.93}$Co$_{0.07}$)$_{3}$V$_{2}$O$_{8}$ (NVCO) is an intermediate compound between NVO and (Ni$_{0.9}$Co$_{0.1}$)$_{3}$V$_{2}$O$_{8}$. The effect of Co$^{2+}$ doping on the system can be revealed by comparing the different behaviors of the three components in high magnetic fields.                                                                                               
	
	The infrared active vibrational mode measurement shows that the phonon in NVO is sensitive to different magnetic states, which leads to a flexible lattice coupling to the spin system\cite{PhysRevB.80.052303}. In CVO, the Co$^{2+}$ displacement plays an important role in local lattice distortion\cite{PhysRevB.81.012403}. Hence the slight changes in the spin system can be captured by a size measurement. Here we present magnetostriction and thermal expansion measurements on the multiferroic NVCO and map out the magnetic phase diagrams as a function of temperature and magnetic field along three crystallographic axes. These results reveal the high field behavior of NVCO. The Co$^{2+}$ doping has a nonlinear effect to the spin correlation of NVO and the 7$\%$ doping leads to the breaking of weak ferromagnetic moment. 
	
	The single crystal was grown by Barilo and Shiryaev \cite{czax3v2o8sample}, and its magnetic structure was described in Ref.\cite{PhysRevB.88.174412}. The crystal was cut into a cuboid along different crystallographic orientations. The magnetostriction and thermal expansion  were measured using a miniature dilatometer\cite{doi:10.1063/1.4748864,mini-dilatometer} and a Andeen Hagerling 2700A along three crystallographic directions, with the magnetic field along three crystallographic directions, respectively. These measurements were conducted in a superconducting magnet with a maximum field up to 16 T and a water cooling magnet with a maximum field up to 35 T in the High Magnetic Field Laboratory of the Chinese Academy of Science. 
	
	\begin{figure}[b]
		\includegraphics[width=8cm]{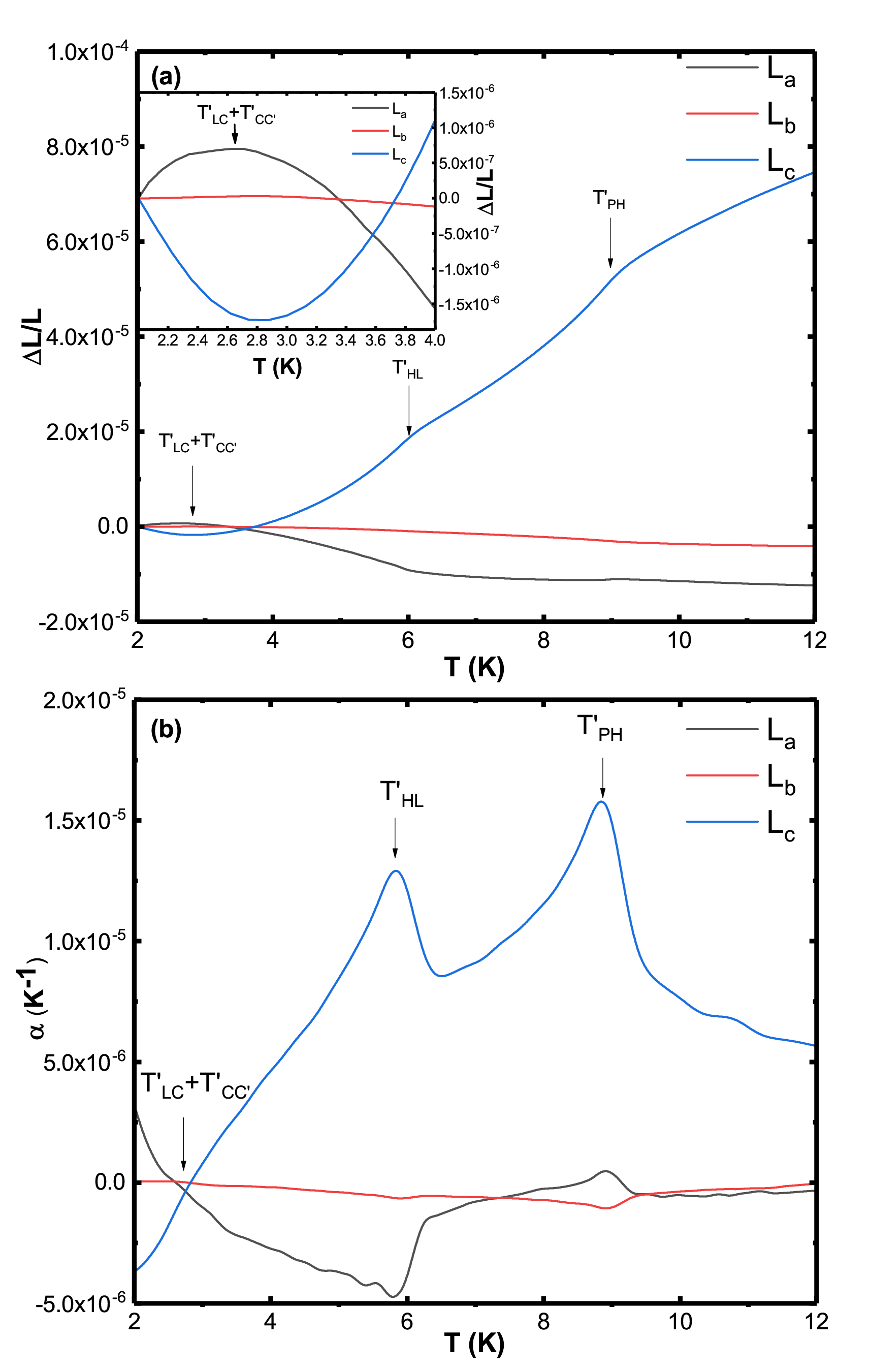}
		\caption{The thermal expansion of NVCO at zero magnetic field along three crystallographic directions at zero field: (a) the linear variation, the inset shows the linear variation from 2 K to 4 K; (b) the expansion rate.}\label{lalblc}
	\end{figure}
	
	The thermal expansion curves of NVCO in zero magnetic field (ZF) along three crystallographic directions are shown in Fig. \ref{lalblc}, including the linear variation and expansion rate. The expansion rate curves show clearly the evolution of phase transitions. At the N\'{e}el temperature (T$^{'}$$_{PH}$=8.9 K), the system undergoes a magnetic transition from the P phase to the long range order HTI phase. As the temperature decreases, the cycloidal LTI phase emerges at T$^{'}$$_{HL}$=5.8 K. These phase transition temperatures are similar to those of NVO, indicating that the short range interactions(NN and NNN interactions) are less affected by the doping Co $^{2+}$. Below T$^{'}$$_{HL}$ the curves exhibit a wide trough and  crest along the $ a $ and $ c $ directions, similar to the electric polarization measurement\cite{PhysRevB.88.174412}. This result indicates a second order transition or continuous transition. In the parent compound NVO, the LTI phase transforms to the C phase at T$_{LC}$ by a first order transition and then goes into thego C$^{'}$ phase at T$_{CC'}$\cite{PhysRevLett.93.247201}. The vanishing of the first order transition and emergence of the second order transition indicate a multicritical point. It means that the multicritical point of T$^{'}$$_{CC'}$ and T$^{'}$$_{LC}$ in NVCO has moved to zero field. For (Ni$_{0.9}$Co$_{0.1}$)$_{3}$V$_{2}$O$_{8}$, the electric polarization measurement indicated that  Co$^{2+}$ could suppress the collinear antiferromagnetic order and lead to the disappearance of theance multicritical point\cite{PhysRevLett.93.247201}. 
	
	\begin{figure*}
		\centering
		\includegraphics[width=18cm]{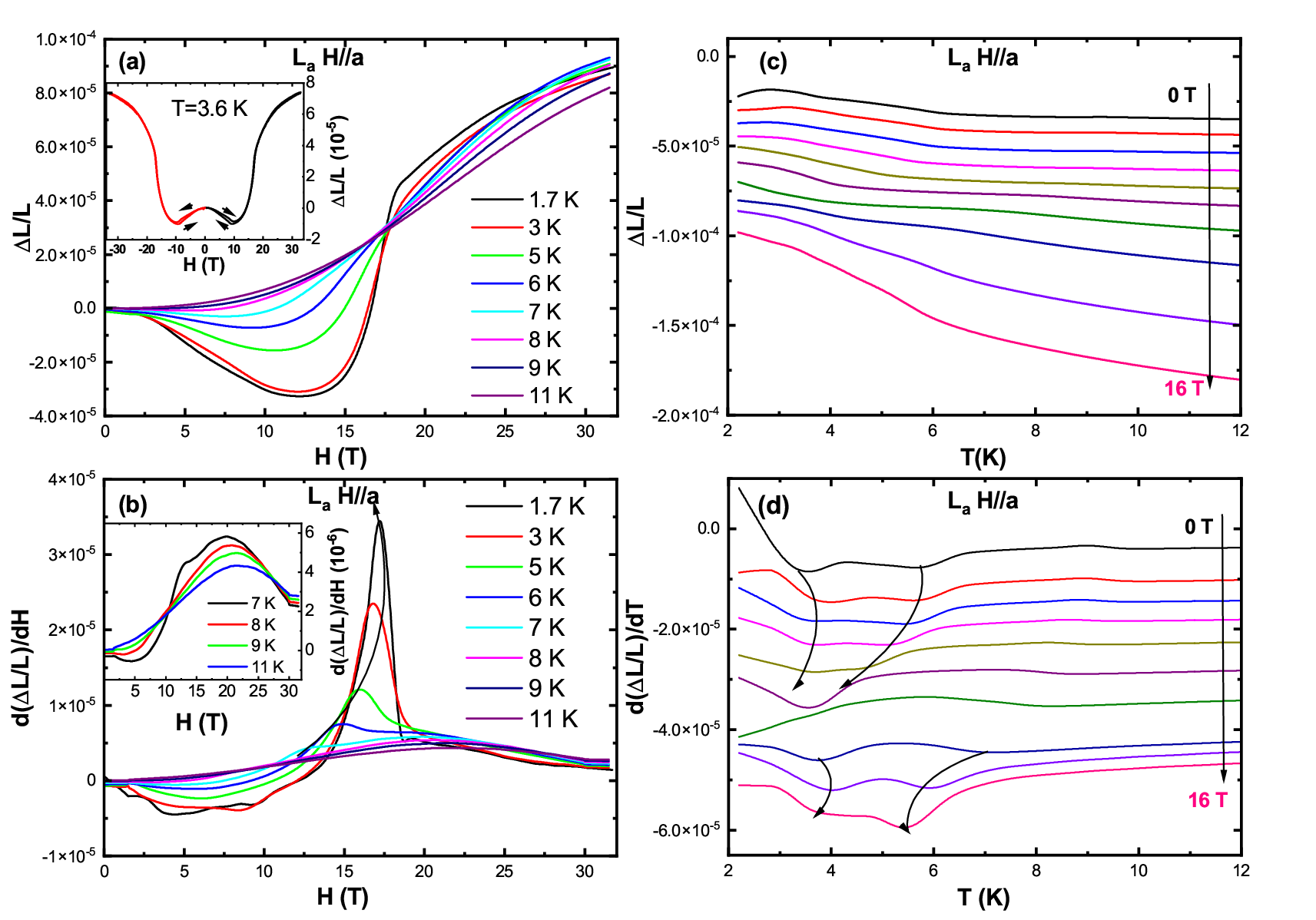}
		\caption{(a) The magnetostriction along the $ a $ direction while H//$ a $, the inset is the magnetostrictive curves with up and down fields at 3.6 K. (b) The differentiation of (a), the inset is the differentiation from 7 K to 11 K. (c) The thermal expansion along the $ a $ direction while H//$ a $. Corresponding to 0,  1, 3, 5, 7, 9, 11, 13 ,15 and 16 T from top to bottom. (d) The differentiation of (c).}\label{hala}	
	\end{figure*}
	The magnetostrictive curves of NVCO along the $ a $ direction are shown in Fig. \ref{hala}(a). The magnetostrictive curves with up and down magnetic fields at 3.6 K are shown in the inset. These curves are symmetric, meaning that the  magnetic field does not influence the magnetostrictive measurement. There are two shapes resembling a continuous phase transition of a valley and a turn, respectively, which means that an intermediate state separates the low field and high field phases. This is a common feature in most anisotropic antiferromagnets. A possible reason is that some terms in the Hamiltonian couple the order parameters of two magnetic phases, leading to the coexistence of two magnetic orders\cite{PhysRevB.11.478}. Here, the intermediate state corresponds to the F2 phase and separates the F1 and F3 phases. It is worth noting that hysteresis appears at about 10 T in the inset. In addition, weak hysteresis appears in the magnetization measurement at 10 T and a distinct hysteresis appears in the volume magnetostriction at 17 T\cite{supplementary}. This is not consistent with the expected second orders transitions. These results imply more than two magnetic order in the F2 phase. As the temperature increases, the valleys and turns are slowly smoothed out. The turning point desisappears above 4 K but the corresponding peak in differentiatiin [Fig. \ref{hala} (b)] disappears until 7 K. In addition, at 7 K a new peak appears. The inset of Fig. \ref{hala} (b) shows that the new peak moves to high fields as the temperature increases. At 1.7 K, the curve tends to flatten out at the higher field, just as the magnetization plateau in NVO\cite{PhysRevB.84.220407}. Fig. \ref{hala} (c) and Fig. \ref{hala} (d) show the thermal expansion of NVCO and its differentiation along the $ a $ direction while applying a specific magnetic field. To distinguish variations between these curves, these curves are shifted in an order that the magnetic field gradually increases from top to bottom. Curved arrows are used to indicate the directions in which the phase transition is moving. When fields are lower than 7 T, the thermal expansions are accompanied by two turning points, corresponding to the LTI-HTI and HTI-P phase transitions, respectively. As the magnetic fields increas, the LTI-HTI and HTI-P phase transitions tend to meet at 9 T, as pointed to by theto two arrows. At 11 T, the thermal expansion exhibits different behavior. Above 11 T, two turning points appear and two arrows are used to mark the evolution of the transitions[Fig. \ref{hala} (d)]. These results imply that the NVCO goes through an intermediate state to high field states.
	
\begin{figure*}
		\centerline{\includegraphics[width=20cm]{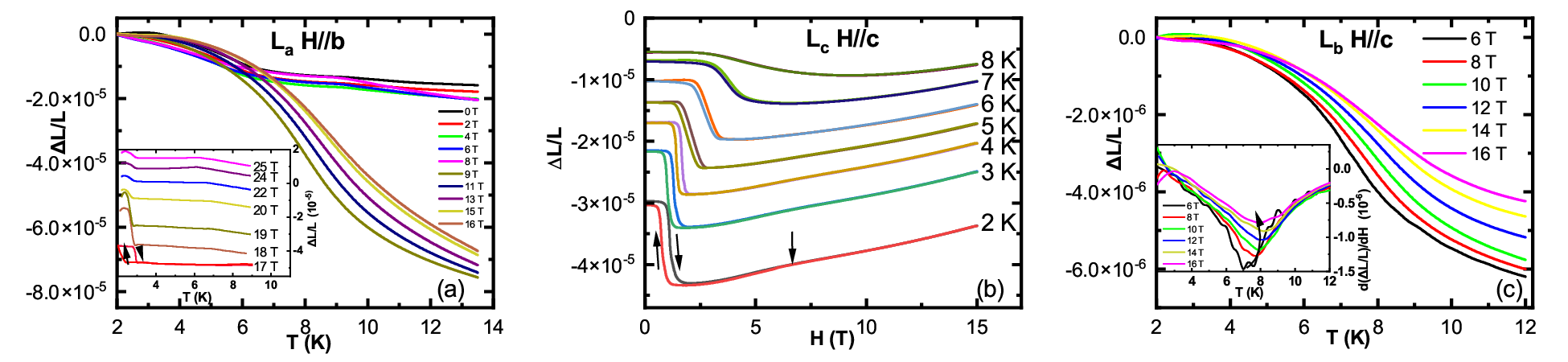}}
		\caption{(a) The thermal expansion along a direction while H//$ b $, the inset is the thermal expansion between 17-25 T. (b) The magnetostriction along the c direction while H//$ c $. (c) The thermal expansion along the $ b $ direction while H//$ c $, the inset is its differentiation.}\label{hbhc}	
	\end{figure*}

	Fig. \ref{hbhc}(a) shows the thermal expansion of NVCO along the $ a $ direction while H//$ b $. Below 8 T, two turning points appear at about 9 and 6 K, corresponding to the phase transitions from the P phase to the HTI phase and then to the LTI phase. At 9 T, different patterns exhibit in the thermal expansion, in which the turning points are smoothed out and the phase transition becomes inconspicuous. Above 17 T, the turning points reappear with a tendency similar to the situation at low field, as shown in the inset of Fig. \ref{hbhc}(a). This phenomenon implies a possible result that the low field states go through intermediate states to the high field states that are analogous to the low field states. Note that a structural phase transition with significant hysteresis is found at 17 T. Theory and experiment have shown that such lattice mutations are possible during magnetic phase transitions\cite{PhysRev.129.578,PhysRevB.77.014407}. It indicates a transition from strong to weak spin lattice coupling.
	
	Fig. \ref{hbhc}(b) exhibits the magnetostriction of NVCO along the $ c $ direction while H//$ c $. We shift curves to make it clearer and use arrows to mark the up and down fields. A sharp jump appears and becomes smooth as the temperature increases. The hysteresis is obvious below 8 K, indicating a first order transition. The hysteresis is also found in the magnetization measurement\cite{supplementary}. Below 6 K, as the magnetic fields increase, the cross-tie spin of the LTI phase is broken by the magnetic fields, resulting in an antiferromagnetic order on the spine site. The abrupt change of the macroscopic crystal sizes also implies a structural difference between the incommensurate phase and commensurate phase. Above 6 K, the system goes into the HTI phase. Different from the transition below 6 K, the structural mutation of the HTI-C transition is weakened although the hystereses remain. A possible reason is that the transition is close to the multicritical point. In addition, an intriguing phenomenon here is that a weak transition follows a sharp jump and disappears above 6 K. It implies an intermediate state between the LTI and C phase. We call it the C$^{''}$ phase. This weak transition also indicates that the C$^{''}$ and C phases are isomorphic. At 8 K, both the hysteresis and a structural mutation disappear, which is a signal of a second order transition. This result shows that above 8 K the HTI phase goes into the P phase as the fields increase. The thermal expansion of NVCO along the $ b $ direction for H//$ c $ is shown in Fig. \ref{hbhc}(c). These curves show a smooth transition between the C and P phases. The transition can be noticed by a differentiation in the inset. The arrows are used to indicate the movement of transition points, indicating the phase transition moves in an arc as the magnetic fields rise. It is worth mentioning that in the parent compound NVO and the other compound, the C-P boundary has not been found explicitly\cite{PhysRevB.74.014429,PhysRevB.94.174441}.
	
\begin{figure*}
		\centerline{\includegraphics[width=20cm,height=7cm]{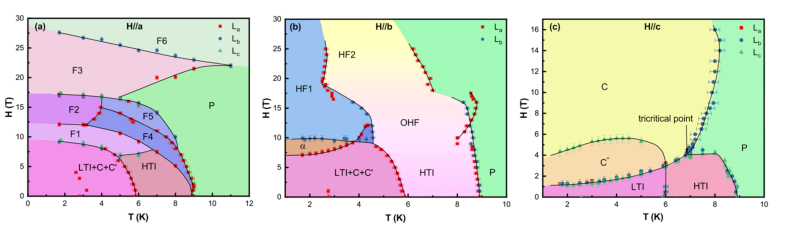}}
		\caption{The magnetic field-temperature diagram along three crystallographic axes. }\label{diagram}	
	\end{figure*}

	In Fig. \ref{diagram}, the magnetic phase diagrams along three crystallographic directions are produced by magnetostriction and thermal expansion data. Based on the Heisenberg nearest-neighbor exchange approximation analysis, in NVO, two subsystems are weakly coupled by a zero mean field of cross-tie sites which is produced by spine sites \cite{PhysRevLett.93.247201,PhysRevB.73.094425}. It leads to the merging of T$_{CC'}$ and T$_{LC}$ in the low magnetic fields. For H//$ a $, this behavior exists in both NVO and CVO\cite{PhysRevLett.93.247201,doi:10.1143/JPSJ.76.034706}. However, in NVCO, the multicritical point appears at 2.7 K and ZF. As the magnetic field increases, the boundary between the C$^{'}$, C and LTI phases becomes blurry(Fig. \ref{diagram}(a)). It implies that the weak coupling is diminished by the doped Co$^{2+}$. The system enters a mixed state with the C$^{'}$, C and LTI phases, which is certified by the electrical polarization data along the $ b $ direction and heat capacity data in Ref. \cite{PhysRevB.88.174412, PhysRevB.84.064447}.  In addition, the multicritical point disappears while doping 10$\%$ Co$^{2+}$\cite{PhysRevB.94.174441}. These results manifest that doped Co$^{2+}$ modulates the spin of the buckled lattice and suppresses the commensurate antiferromagnetic phase. In terms of the Zeeman energy of the system, the spin moment of the C phase is parallel to magnetic field for H//$ a $, leading to a minimum Zeeman gain. It exacerbates the instability of the C phase\cite{PhysRevB.74.014429}. 
	In the diagram for H//$ a $ [Fig. \ref{diagram}(a)], the high field phase (F3) and low field phase (F1) are separated by the F2 phase, similar to the parent compound. This is a common feature of anisotropic antiferromagnets\cite{PhysRevB.84.220407,PhysRevB.11.478}. But the field dependence stabilities of the LTI and F2 phases are improved in the doped sample. Above the HTI phase, there are two intermediate phases which have anisotropy in the magnetostriction, F4 and F5. The F4 phase can be detected in the magnetostrictive measurement along the $ a $ direction and the F5 phase can be detected along the $ b $ direction. According to the phase diagram, we found that the F3 phase is abnormal. As the field increases, the F3-P transition moves to a higher temperature, which indicates that the F3 phase is a field-induced phase. The inset of Fig. \ref{hala} (b) shows this tendency by the differentiation in magnetostriction. In the higher field, the high field phase(F6) maintains similar behavior with the parent compound\cite{PhysRevB.84.220407}. More data for the boundary evolution of F6 are provided in Ref.\cite{supplementary}.
	
	In Fig. \ref{diagram}(b), the temperature and magnetic field phase diagram for H//$ b $ is built up. The boundary between the C$^{'}$, C and LTI phases is still unclear. An unknown phase between the LTI and high field phase (HF1) is found, which seem to be the $\alpha$ phase in Ref.\cite{hbdiagram1}. Here, the clear up and down boundaries are built by two first-order transitions. There is currently no more information about this phase and further experiments are needed. The phase transitions between HTI, OHF (the optical high field phase between HTI and HF2 as reported in a magneto-optical measurement\cite{PhysRevB.74.235101}) and HF2 phases are not found in the magnetostrictive measurement. However, the diversity in the thermal expansion measurement from 18 to 19 T (Fig. \ref{hbhc}(a)) implies the existence of OHF. Note that two interleaved boundaries between the OHF and P phases are obtained by the data along the $ a $ and $ b $ directions, respectively. These results imply that the OHF phase could be mixed state of two anisotropic phases, which would explain the blurry boundaries between HTI, OHF and HF2 phases.  The diagram for H//$ c $(Fig. \ref{diagram}(c)) is simply relative. As Fig. \ref{hbhc}(b) shows, there is an intermediate phase between the LTI and C phases, which we called the C$^{''}$ phase. Since this transition disappears above 6 K, it is possible to meet with the LTI at 6 K. It is worth noting that this weak transition is only detected along the $ c $ direction, which means that the C and C$^{''}$ phases are possibly isomorphic. We find that the LTI-C$^{''}$ phase boundary changes in an exponential law by intensive measurements. Moreover, the distinct C-P transition arises in the thermal expansion measurement along the $ b $ direction. The C-P boundary indicates a distinct field induced effect, close to that of NVO and (Ni$_{0.9}$Co$_{0.1}$)$_{3}$V$_{2}$O$_{8}$\cite{PhysRevB.74.014429,PhysRevB.94.174441}. 
	
	In NVO, considering the DM interaction, the spine staggered moment N$_{s,a}$ generates a weak ferromagnetic (FM) moment M$_{s,c}$ in the spine site along the $ c $ direction:\cite{PhysRevB.74.014429} 
	\begin{equation}
		M_{s,c}\approx2D_b\chi_{s,c}(0)N_{s,a}.  \label{eq:1}
	\end{equation}
	where the D$_{b}$ is the $ b $ component of DM vector, $\chi$$_{s,c}$(0) is the $ c $ component of spine susceptibility with zero wave vector. Analogously, on the cross-tie sites a FM moment will generate from the so-called off-diagonal exchange spine cross-tie interaction:
	\begin{equation}
		M_{c,c}=-4\chi_{c,c}(j_{ac}-d_b)N_{s,a}. \label{eq:2}
	\end{equation}
	where the $\chi$$_{c,c}$ is the $ c $ component of the cross-tie susceptibility, j$_{ac}$ and d$_{b}$ are the symmetric (PD) and antisymmetric (DM) elements, respectively. M$_{s,c}$ and M$_{c,c}$ form the M$_{c}$, the weak FM moment along the $ c $ direction, which leads to the linear boundary in the HTI-C and LTI-C transitions while H//$ c $. The bilinear coupling between M$_{c}$ and the spine staggered moment N$_{s,a}$ gives rise to a blurry boundary between the P and C phases. If this coupling does not exist, a tricritical point with two second-order lines (P-HTI and P-C) and a first-order line (HTI-C) will appear, which is the result of a competition between the HTI and C phase. We find that two second order transitions(P-HTI and P-C) and a first order transition(HTI-C) as well as a tricritical point appear in the phase diagram [Fig. \ref{diagram}(c)] while doping 7$\%$ Co$^{2+}$ on NVO. The linear boundary between the incommensurate and commensurate phases disappears and an exponential LTI-C$^{''}$ boundary and nonlinear HTI-C boundary appear. This result indicates that the M$_{c}$ and N$_{s,a}$ have decoupled. Doping Co$^{2+}$ changes the spin orientation of the spine and cross-tie sites. It affects the PD and DM interaction and then breaks the weak FM. Note that the (Ni$_{0.9}$Co$_{0.1}$)$_{3}$V$_{2}$O$_{8}$ maintains similar behavior with NVO\cite{PhysRevB.94.174441}, showing obviously that the effect of doping Co$^{2+}$ on NVO is nonlinear.   
	
	We now discuss these behaviors of the phase boundary by a symmetry analysis. In the NVO, the P phase has double rotation and inversion symmetry with the representation $\varGamma$$_{7}$ for H//$ c $ and $\varGamma$$_{5}$ for H//$ a $. But the C phase has  representation $\varGamma$$_{7}$ both for H//$ a $ and H//$ c $\cite{PhysRevB.74.014429}. These symmetry relations result in that the C phase only transforms from the LTI and HTI phase into the P phase for H//$ a $, and it goes into the P phase without a visible transition for H//$ c $. Based on our experimental observation, for H//$ c $, the C phase goes into the P phase by a second order transition and the C-P phase boundary appears. It indicates the symmetry of the C phase in NVCO is different from the parent compound.
	
	In summary, we have provided detail investigations for the magnetic phase diagrams of NVCO by magnetostriction and thermal expansion measurements. The behavior of NVCO in high magnetic fields and the effect of doping Co$^{2+}$ on the NVO system are discussed. For H//$ a $, the multicritical point of the C$^{'}$, C and LTI phase moves to ZF and the boundary between them becomes blurry, indicating that the commensurate phase is suppressed by Co$^{2+}$. The field dependence stability of the LTI phase is improved distinctly. As the magnetic field increases, the system goes into the high field states (F3 and F6). The F3 phase exhibits a field induced behavior. In addition, there are two intermediate phases (F4 and F5) with distinct anisotropy between the HTI and P phases. For H//b, a mixed state of the OHF phase consisting of two anisotropic phases is proposed. At about 20 T and 3 K, the system undergoes a transition from strong to weak spin lattice coupling. The high field states share a similar evolution with the low field states. In addition, the diagram reveals an unknown phase that is similar with the $\alpha$ phase. It is separated by two first order transitions. For H//$ c $, an intermediate phase (C$^{''}$) between the LTI and C phase is detected, which may be isomorphic with the C phase. The nonlinear LTI-C$^{''}$ and HTI-C phase boundaries indicate the breaking of weak FM on C phase. The clear C-P phase boundary indicates the decoupling between magnetization M$_{c}$ and the spine staggered moment N$_{s,a}$. Compared with the NVO and (Ni$_{0.9}$Co$_{0.1}$)$_{3}$V$_{2}$O$_{8}$, we find that doping Co$^{2+}$ produces a nonlinear effect on the NVO system. With the increase of Co$^{2+}$ doping, the weak FM moment first disappears and then reappears. These results manifest that a small amount of Co$^{2+}$ doping can effectively adjust the spin interaction between the competing magnetic orders in NVO system. 
	
	\begin{acknowledgments}
		We are grateful to Vassil. Skumryev and Alexander Mukhin for their critical reading of the manuscript and suggestions. This work was supported by the National Natural Science Foundation of China (Grants No. 11874359, No. 11704385) and Systematic Fundamental Research Program Leveraging Major Scientific and Technological Infrastructure, Chinese Academy of Sciences under contract No. JZHKYPT-2021-08. A portion of this work was performed on the Steady High Magnetic Field Facilities, High Magnetic Field Laboratory, Chinese Academy of Sciences, and supported by the High Magnetic Field Laboratory of Anhui Province. 
	\end{acknowledgments}

\end{document}